\documentclass[12pt,preprint]{aastex}

\def\etal{\rm{et al. }}
\def\deg{^\circ }

\tighten
\begin{document}
\title{\bf {\it Hubble Space Telescope} Imaging of Bipolar Nuclear Shells 
in the Disturbed Virgo Cluster Galaxy NGC~4438\footnote{
Based on observations with the {\it Hubble Space Telescope} which is
operated by AURA, Inc., under NASA contract NAS 5-26555}
}
\author {Jeffrey D. P. Kenney\altaffilmark{2} \& Elizabeth E. Yale \altaffilmark{2}}
\email{kenney@astro.yale.edu, eey2@astro.yale.edu}
\altaffiltext{2}{Yale University Astronomy Department, P.O. Box 208101, New Haven, CT
06520-8101}

\received{May 3, 2000}
\revised{October 10, 2001}

\shorttitle{Nuclear Outflow Shells in NGC~4438}
\shortauthors{Kenney \& Yale}
\begin{abstract} 
We present broadband and narrowband
{\it Hubble Space Telescope} images of the 
central region of the heavily disturbed Virgo cluster galaxy NGC~4438 (Arp120),
whose nucleus has been described as a type 1 LINER or dwarf Seyfert.
Narrowband H$\alpha$ and [NII] HST images reveal striking bipolar shell 
features, 1 kpc in projected length from end-to-end,
which are likely the result of an outflow from the nuclear region
experiencing a strong interaction with the ISM.
While these outflow shells share similarities with those in some other
starburst or AGN galaxies,
these in NGC~4438 are notable because NGC~4438 harbors neither a 
luminous circumnuclear starburst nor a luminous AGN.

The shells appear to be closed at their outer ends, suggesting that the outflow in
NGC~4438 is dynamically younger than those in some other galaxies.
The radio continuum emission is strongly enhanced
near the outer ends of the shells, suggesting working surfaces arising from
collimated nuclear outflows which have impacted and shocked the surrounding ISM.
The 2 shells are quite different, as the
northwestern (NW) shell is luminous and compact, while the
southeastern (SE) shell is 2.5 times longer and much fainter, 
in both optical emission lines and the non-thermal radio continuum.
The differences between the 2 shells may be attributed to a difference in ISM density
on the 2 sides of the nuclear disk. Such an ISM asymmetry exists on larger scales
in this heavily disturbed galaxy.

At the base of the outflow is a nuclear source, 
which is the highest surface brightness source in the galaxy at optical wavelengths,
This source is resolved with a FWHM=0.3$''$=25 pc and
has modest luminosities, uncorrected for extinction,
of 5$\times$10$^{38}$ erg s$^{-1}$ in H$\alpha$,
and M$_{\rm B}$=-13.
We discuss whether the outflow is powered by a low luminosity AGN
or a compact nuclear starburst.
The kinetic energy associated with the ionized gas in the shells is $\sim$10$^{53}$ ergs, 
which could be produced either by massive star formation or an AGN.
While the NW shell, which contributes most of 
the flux in most ground-based apertures centered on the nucleus,
exhibits LINER-type line ratios,
the nucleus has an H$\alpha$/[NII] ratio consistent with an HII region.
Although there appears to be
very little massive star formation occuring in the central kpc,
the nucleus may contain  a partially obscured, young nuclear star cluster.
On the other hand,
the collimation of the southeastern shell, and the strongly enhanced radio
continuum emission at the outer ends of the shells  are more easily
explained by jets associated with a nuclear black hole than a compact nuclear starburst,
although there is no direct evidence of a jet.
The 2000 km s$^{-1}$ broad line component could be due to an AGN
broad line region but might also be due to high velocity gas in the outflow.
Since NGC~4438 has a large bulge, a large nuclear black hole might be expected.
\end{abstract}

\section {Introduction}

The nuclei of galaxies exhibit an enormous range of activity,
and the factors which govern their differences are far from understood.
LINERs (low ionization nuclear emission regions),
which are among the most common type of galactic nucleus,
are clearly a heterogenous group of objects,
and different galaxies likely have different origins for the
excitation of ionized gas whose spectra lead to the LINER classification.
The excitation mechanisms include
photoionization from low luminosity AGN,
photoionization from hot stars,
and shocks from AGN jets or starburst outflows 
(for a review see Filippenko 1996).
Progress towards understanding the nature of LINERs
must involve high resolution studies of nearby nuclei with 
a wide range of properties. 
In this paper we focus on one nearby galaxy 
whose nucleus, classified variously as a LINER 1
or a 'dwarf Seyfert' (due to its nuclear broad line component), 
can reveal some of the physical 
processes which occur in galactic nuclei.

NGC~4438 (Arp 120) is a highly inclined and very peculiar large 
spiral galaxy near the center of the Virgo Cluster (D=16 Mpc). 
Its disturbed stellar distribution indicates 
that it recently experienced some type of major gravitational encounter
(Combes \etal 1988; Moore \etal 1996; Katsiyannis \etal 1998).
Its complex and strongly disturbed
ISM (Kotanyi \etal 1983; Warmels 1988; Cayatte \etal 1990; Keel \& Wehrle 1993;
Kenney \etal 1995)
is quite different from that of any other galaxy presently known,
and it is likely that a high-velocity ISM-ISM collision (Kenney \etal 1995) 
and/or an ICM-ISM interaction (Kotanyi \etal 1983; Keel \& Wehrle 1993) is 
responsible for the ISM disturbance.
While the origin of the disturbance is not a focus of the present 
paper,
it is relevant that much of the ISM is significantly displaced 
to the west of the galaxy's main disk. 
The galaxy disturbance may well be responsible for 
some of the properties of the nuclear region.

The nuclear region of NGC~4438 has a complex and unusual radio continuum morphology 
which has not previously been understood (Hummel \& Saikia 1991; Keel \& Wehrle 1993).
The optical ``nucleus?'' of NGC~4438 has been classified as 
a LINER  (Heckman 1980; Stauffer 1982; Keel 1983),
but as shown here, the source with LINER-type line ratios is not the real nucleus.
The detection of high-velocity gas suggests that the nucleus is 'active'.
Ho \etal (1997) 
find a weak broad component in the H$\alpha$ line whose
FWHM is 2050 km s$^{-1}$, making NGC~4438 a ``dwarf Seyfert'',
or LINER 1.9 nucleus.

In this paper, we present and discuss the results of HST imaging of
the central region of NGC~4438, which reveal
asymmetric bipolar shell features strongly suggestive of
outflow from the nucleus.
The dominant source of both optical line 
and radio continuum emission in the central few
arcseconds (i.e. at ground-based resolution)
of NGC~4438 is not the nucleus itself, but
emission from an outflow which is experiencing a strong
interaction with the surrounding ISM.
We also detect a compact nuclear source (FWHM=~0.3$''$), 
which does not have LINER-type line ratios,
but instead has an H$\alpha$/[NII] ratio similar to an HII region.
We discuss whether the nuclear outflow 
is powered by a compact nuclear starburst or
by jets from a low luminosity AGN.

\section {Observations}

\begin{figure}
\epsscale{0.95}
\plotone{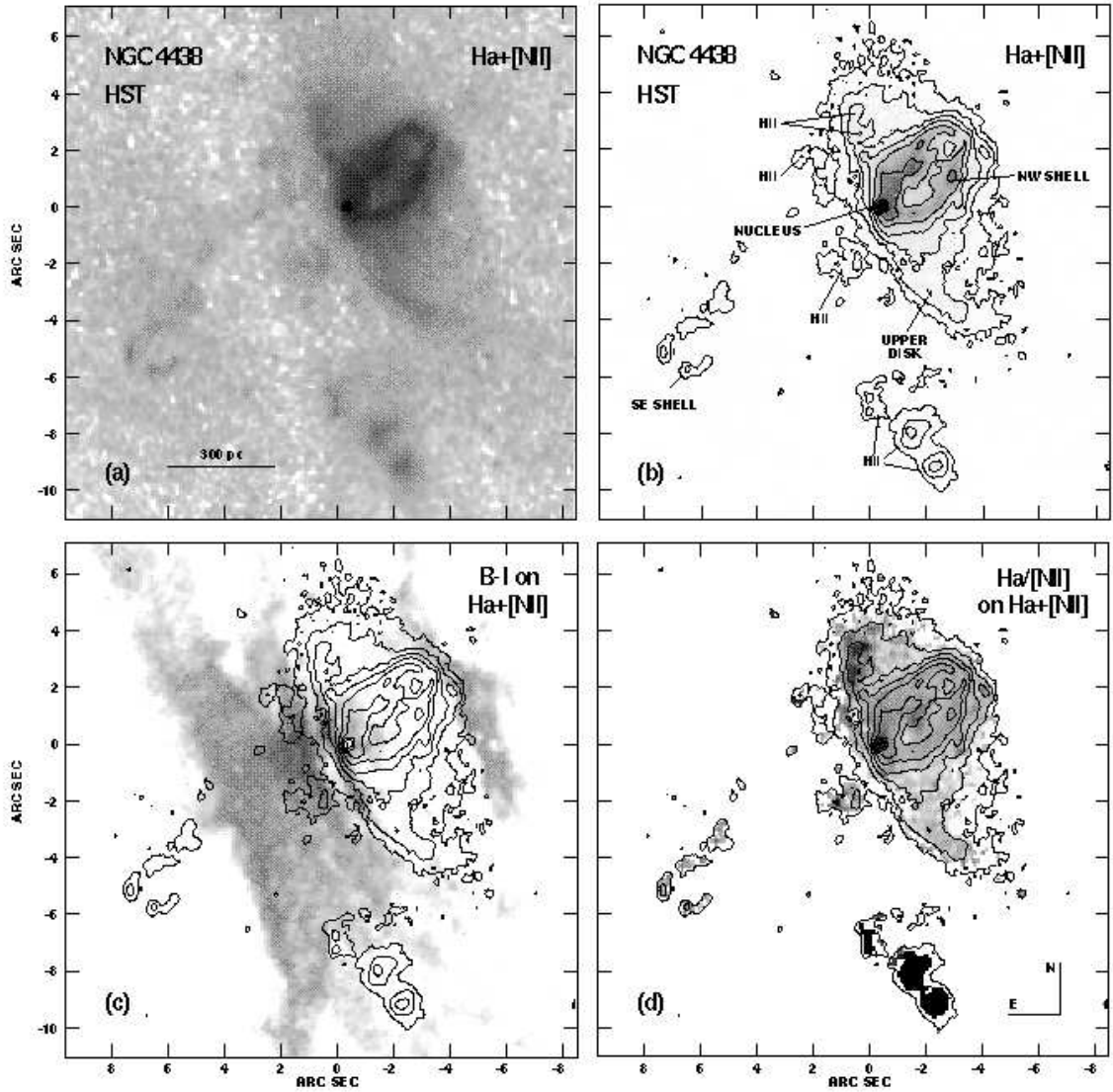}
\caption{a.) Sum of HST H$\alpha$ and [NII] line 
images over central 18$''$ of NGC~4438. Greyscale stretch is logarithmic.
b.) Greyscale and contour maps of H$\alpha$+[NII], with components labelled.
c.) F450W-F814W (B-I) greyscale color map
superposed on H$\alpha$+[NII] contour map.
The gray regions have B-I greater than 1.5,
and are largely associated with dust lanes.
This shows that part of the SE lobe is likely obscured by dust,
and that the line-emitting arc is spatially coincident with the edge
of a dust lane.
d.) H$\alpha$/[NII] ratio greyscale map
superposed on H$\alpha$+[NII] contour map.
The lightest grayscale colors correspond to an  H$\alpha$/[NII] ratio of
0.3, and the darkest colors correspond to an H$\alpha$/[NII] ratio $\geq$2.
The nucleus has a ratio of 1.6, and the luminous HII complex near the
bottom of the frame has ratios between 2-3.
}
\end{figure}

WFPC2 images of NGC4438 were obtained in March 1999
with broadband F450W (B), F675W (R) and 
F814W (I) and narrowband F656N and F658N filters.  The bandpasses 
of F656N and F658N cover the $\lambda$6563\AA\
H$\alpha$ and the $\lambda$6583\AA\ [N~II] lines, respectively, 
and are narrow enough (22.0 and 28.5 \AA\ )
to separate these nearby lines.
Since NGC~4438 has a redshift of only 71 km s$^{-1}$
(Kenney \etal 1995), these emission lines fall within the
bandpasses of these HST filters.
Because of difficulty in finding guide stars,  the nucleus
of the galaxy is located in WF3, which has
a pixel scale of 0.1$''$, corresponding to 8 pc.
At least 3 exposures were made for each of the broadband filters,
with total exposure times of 1300 seconds in the
B filter, 1450 seconds in the  R filter, and 1050 seconds in the 
I filter. The central few pixels were saturated in the I and R images.
Four exposures were made for each of the narrowband filters,
with total exposure times of 5200 seconds for each filter.

All images were processed in the standard HST pipeline using the best available
calibration and reference files.    The IRAF/STSDAS  task  gcombine was used to
remove cosmic rays and other artifacts
from the images and combine the images taken in each filter. 
The task WFIXUP interpolated over bad pixels, and
the task WMOSAIC in IRAF/STSDAS was used to
make mosaic images.
After cosmic rays were removed, continuum-subtracted H$\alpha$ and [N II]
images were made by  subtracting scaled versions of the R image
from the narrowband images.  
For the central few pixels which were saturated in the R image, we instead used
the unsaturated B image. This introduces an additional error in the 
central few pixels of the continuum-subtracted 
H$\alpha$ and [N II] maps, although this error is fairly small
because the line emission is relatively strong here.
Figure 1a shows the sum of the H$\alpha$ and [N II] images in the central 18$''$.

\section {Results}

We use the morphological information shown in the H$\alpha$+[N~II] map of Figure 1a,
the H$\alpha$/[N~II] line ratio map shown in Figure 1d,
and the broadband B-I color map shown in Figure 1c,
to identify 4 types of line emitting components
in the central region of NGC~4438: the nuclear source,
2 outflow shells, a few HII regions, and an arc of diffuse gas
which likely arises from the upper layers of a dusty gas disk.
These components are labelled on the contour plus greyscale
H$\alpha$+[N~II] map in Figure 1b.
A color-coded combination of all 3 broadband and 2 narrowband images,
made by Zolt Levay of STScI, is shown in Figure 2.

\begin{figure}
\plotone{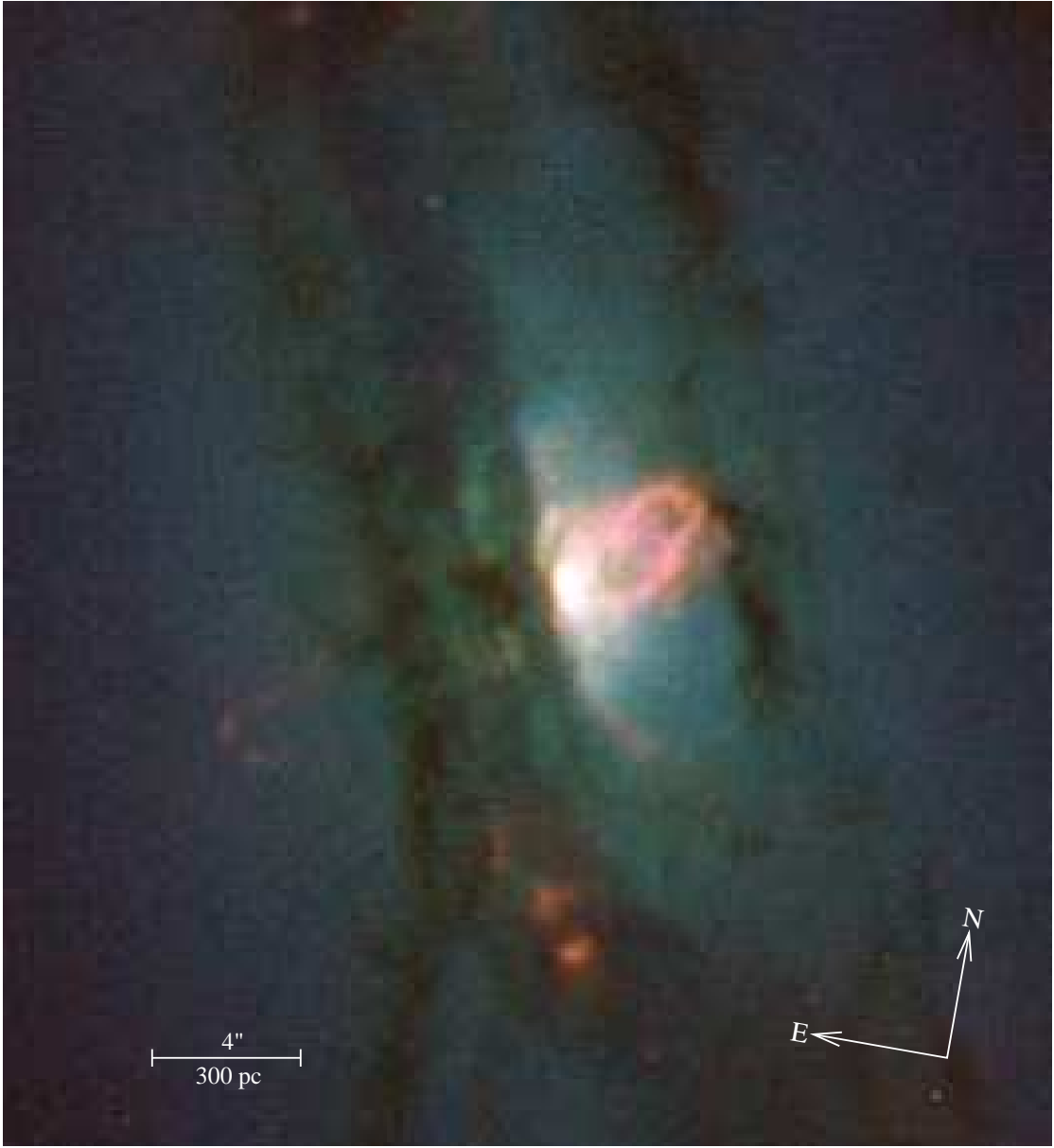}
\caption{
Pseudo-color multiband image of central 20$''$ of NGC~4438,
from HST WFPC2. Color coding is: 
I=red R=green B=blue H$\alpha$=red/magenta, [NII]=red/orange.
The association of arcs of ionized gas with the edges of dust lanes is
particularly evident in this image.
}
\end{figure}

\subsection{Outflow Shells}

The emission line images shown in Figure 1 reveal
a double shell morphology straddling a strong compact source
which is within 1$''$ of the nucleus.
The 2 shells
are quite different, as the northwestern (NW) lobe is bright and compact 
while the southeastern (SE) lobe is elongated and faint.
The NW bubble$\footnote{
Keel \& Wehrle (1993) spoke of an outflow loop or bubble toward the NW,
but this was a much larger and very different feature which 
Kenney \etal (1995) identify as part of the extended filamentary
complex associated with an ISM-ISM or ISM-ICM interaction,
and not assocated with a nuclear outflow. See Figure 5.}$
appears as a complete, edge-brightened, roughly elliptical shell,
although with significant substructure.
The maximum projected radial extent of this shell is 3.8$''$=290 pc.
The nuclear source lies within this elliptical shell and close to
the major axis of the ellipse.
The nuclear source is thus consistent with being the origin of the
energetic outflow presumably responsible for the double bubble morphology. 

The position angle of the H$\alpha$+[NII] shells and the radio continuum
structure (Hummel \& Saikia 1991) is 125$\pm$1$\deg$, which is offset by 6$\pm$2$\deg$
from the minor axis of the galaxy, as determined from
the position angle of the major axis of 29$\pm$2$\deg$ for
the CO disk and the stellar bulge (Kenney \etal 1995).
Thus the bipolar outflow appears to be oriented 
a bit offset from perpendicular 
to the stellar disk
and to a disturbed circumnuclear ring of molecular gas and dust $\sim$1 kpc = 13$''$ 
in radius.
The B-I map in Fig. 1c shows a red feature, probably a dust lane, located in projection
just above the top of the NW shell, suggesting that a region of enhanced gas density
may play a role in determining the spatial extent of the NW shell.

Figure 1d reveals that the H$\alpha$/[NII] line ratios 
are 0.7-1.0 in the brightest regions of the NW shell,
and 0.5-0.7 in the lowest surface brightness regions in the shell interior.
It is this bright shell emission which is responsible for the LINER 
line ratios measured in ground-based spectroscopy (Stauffer 1982; Keel 1983;
Ho \etal 1997), and not the nucleus itself (see below).
LINERs are a heterogenous class.
Some nuclei have LINER line ratios on $<$10 pc scales,
and for such sources the LINER line ratios may arise from photoionization
if there is a radial gradient in the gas density (e.g. NGC~4579, Barth \etal 2001).
However, in NGC~4438, the shell LINER emission likely arises from shock excitation.

\begin{figure}
\plotone{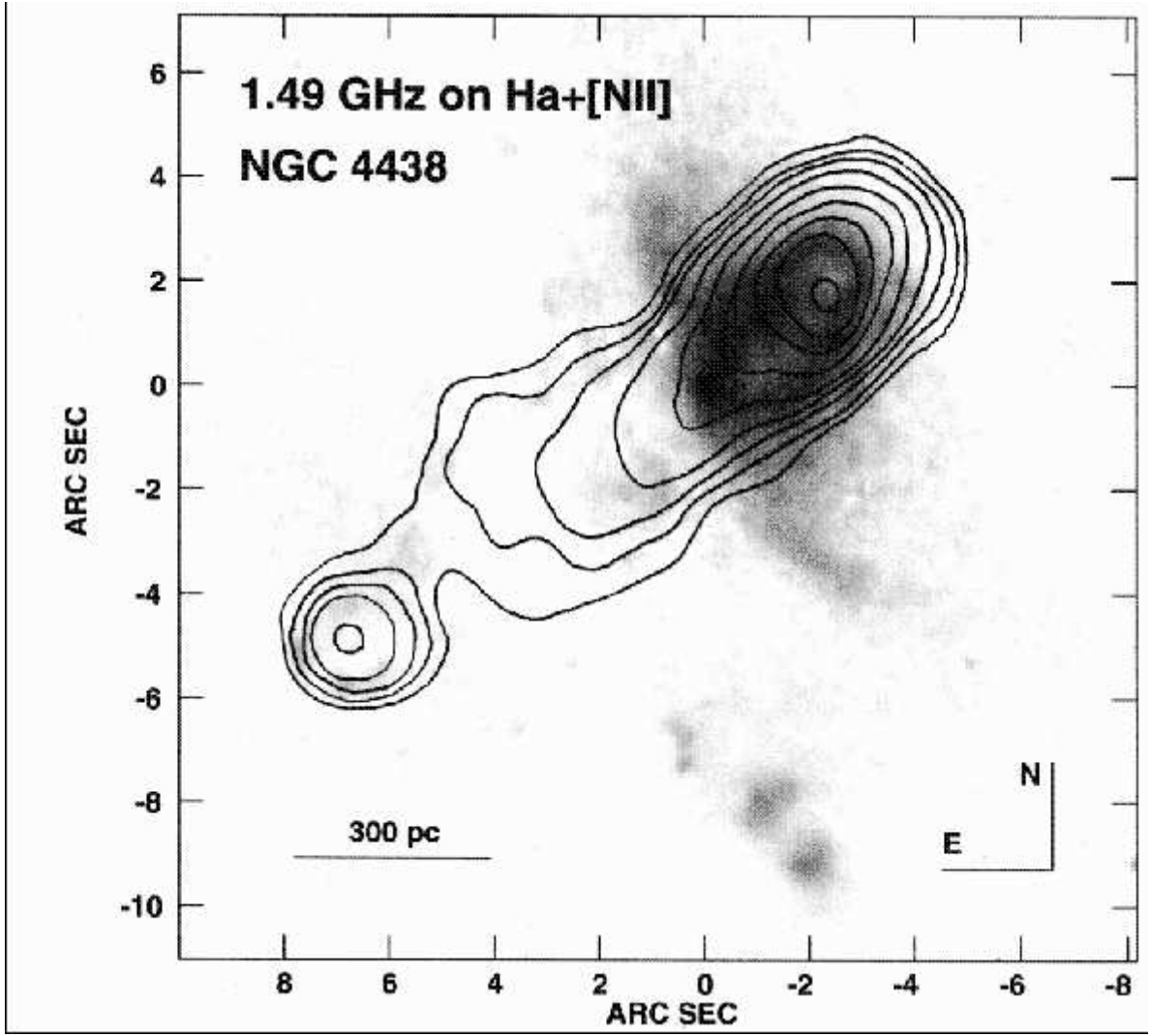}
\caption{
HST H$\alpha$+[NII] greyscale image of NGC~4438,
superposed on 1.49 GHz radio continuum contour map at 1.2$''$ resolution
from Hummel \& Saikia (1991).
Note that the NW shell is much brighter and more compact 
than the SE shell in the non-thermal radio continuum as well as in H$\alpha$+[NII].
Note also the enhanced radio continuum emission near the outer ends of the shells,
suggesting working surfaces. Greyscale stretch is logarithmic, and
contours are 0.2, 0.45, 0.75, 1.25, 2.5, 5, 7.5, 10, 12.5 mJy beam$^{-1}$.
}
\end{figure}

Unlike the NW shell, optical line emission from the SE shell cannot
be easily traced to the nucleus.
Much of the SE shell is likely obscured by dust. Indeed, the broadband 
HST images show several dust lanes in this region, and the B-I map
in Figures 1c and 2 shows that a major dust lane lies between the nucleus
and the visible part of the SE shell.
The 2 lobes are also intrinsically quite different.
The outer radius of the SE shell is 2.5 times further from the
nucleus (9.5$''$=732 pc) than the corresponding measure for the NW shell,
and this cannot be the result of extinction.
The opening angle of the SE feature is also significantly less than
that of the NW shell.

\begin{figure}
\plotone{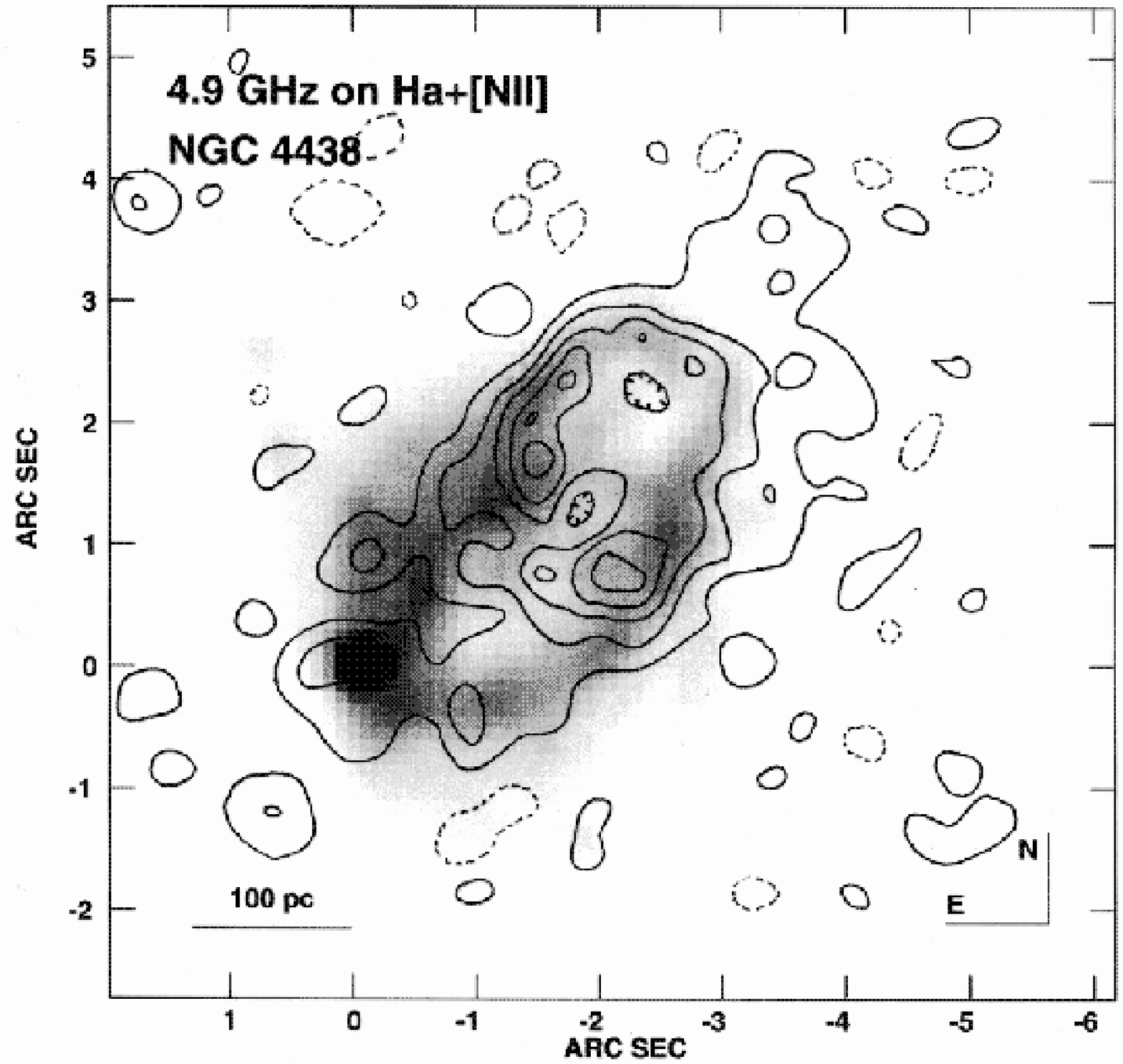}
\caption{
HST H$\alpha$+[NII]  greyscale image of NGC~4438,
superposed on 4.86 GHz radio continuum  contour map 
at 0.4$''$ resolution
from Hummel \& Saikia (1991).
Note that compared to the H$\alpha$+[NII] emission,
the radio continuum emission is relatively stronger inside the shell,
and near the outer end of the shell.
Note also the  presence of radio continuum emission from the nucleus.
Greyscale stretch is linear, and
contours are -0.1, 0.1, 0.3, 0.5, 0.7, 0.9 mJy beam$^{-1}$.
}
\end{figure}

Moreover, the NW shell is much brighter than the SE shell
in the radio continuum as well as in optical line emission (Hummel \& Saikia 1991).
Figures 3 and 4 show the HST H$\alpha$+[NII] image superposed on the
radio continuum maps of Hummel \& Saikia (1991).
The 1.4$''$ resolution 1.4 GHz VLA map in Figure 3 shows a good overall
correspondance on large scales
between the HST H$\alpha$+[NII] and the radio continuum emission.
In particular, the SE radio continuum component
is 2.5 longer and 15 times fainter than the western component.
Thus the SE and NW components are intrinsically different,
and the faintness of the eastern shell in the optical is not entirely 
due to dust extinction.
The 0.4$''$ resolution 4.86 GHz VLA map in Figure 4 shows that the luminous
NE source has a shell-like morphology with sub-structure in the radio continuum as 
well as in the H$\alpha$+[NII] line. This map
also shows a weak  radio continuum nuclear source.

Although there is a rough spatial correlation of radio continuum and
line emission,  there are significant differences.
Whereas the H$\alpha$ and [NII] surface brightness
is similar in the near-nuclear and outer portions of the shell,
the outer shell has much stronger radio continuum emission.
This suggests working surfaces in which 
collimated high-velocity nuclear jets or outflows have impacted and shocked
the surrounding ISM.
As compared to the outer edges of the shell, the interior of the shell
has a greater relative brightness in  radio continuum than H$\alpha$+[NII],
suggesting that the optical line emission only arises at the interface
between the outflowing material and the ISM, whereas the radio continuum
emission arises from the interior as well as the interface.

This close correspondance between optical narrow line region and 
radio continuum emitting regions is
similar to what is observed in many Seyfert galaxies and other AGN.
In many such cases it appears that
NLRs form cocoons around the radio jets, probably because
shocks driven into the ISM by a jet power the narrow line region.
(e.g., Capetti \etal 1999).

\subsubsection{Substructure of NW Shell}

There is substructure in the NW shell, seen especially in the 
radio continuum and optical overlay of Figure 4. 
Morphologically the NW shell seems to have 2 components, and inner
and an outer shell, each extending about half the total length.
The ratio of radio continuum to H$\alpha$ surface brightness
is much higher in the outer than the inner shell.
The ridges of radio continuum curve around and partly cross the interior of
the shell, nearly forming a secondary bubble within the larger
NW bubble. 
It appears that
the NW bubble may have partially ruptured, and plasma containing
most of the relativistic electrons has escaped the initial bubble
and formed a secondary bubble.
This is similar to NGC~3079, in that the inner part of the outflow
has relatively strong  H$\alpha$ emisison, whereas the outer part has
relatively strong radio continuum  emission (Veilleux et al. 1994).

The mid-axis of the outer shell is not aligned with
the mid-axis of the inner shell, but is displaced toward the north.
The offset outer shell gives the
entire NW shell the appearance of being skewed, but
actually the mid-axis of the inner shell lies close to the line connecting the 
SE shell and the nucleus.
It may be that the breakout point of the secondary bubble
is not at the apex of the primary bubble, but offset to the N of the apex.

The asymmetry of the NW and SE shells is discussed in \S 4.3.

\subsection{HII Regions}

Although there is substantial optical line emission from NGC~4438,
the H$\alpha$/[NII] ratio from most if it is less than 1, indicating that 
photoionization by hot stars is not the dominant source of gas excitation
(Keel 1983, Keel \& Wehrle 1993).
This is true in the central region, as well as 
globally (Kenney \etal in prep).
A modest number of HII regions have been detected in the HST line images.
These can be easily identifed by their compact morphology, and by
relatively high localized H$\alpha$/[NII] line ratios which are between
1.3 and 3.0, as shown in Fig. 1d. The HII regions which we have identified 
in the central region are indicated in Figure 1b.

Excluding the nucleus itself,
there are no large HII regions apparent within $\sim$1$''$ of the nucleus,
as judged by the low values of the H$\alpha$/[NII] line ratio. 
While there is some massive star formation occuring in the central few
hundred pc of NGC~4438, the star formation rate is quite low.
This galaxy is not experiencing a circumnuclear starburst,
by which we mean vigorous star forming activity over a region
of a few hundred pc, but could be experiencing or have experienced
a more compact nuclear starburst.

The brightest star-forming complex in the galaxy is located
8$''$=600 pc (projected) south of the nucleus, apparently near the
outer edge of the circumnuclear disk of gas and dust,
and near the bottom of Figure 1.
The measured  H$\alpha$+[NII] flux of the entire complex is
2.0$\times$10$^{-14}$ erg s$^{-1}$ cm$^{-2}$, with most of the flux
arising from 2 components which are $\sim$1$''$=77 pc apart.
The corresponding  H$\alpha$+[NII] luminosity of 
6$\times$10$^{38}$ erg s$^{-1}$ is about an order of magnitude
less than 30 Doradus in the LMC, and about two orders of magnitude
greater than the Orion Nebula.

\subsection{Arc}

There is an arc of line emission which passes close to the nucleus
and curves toward the west,
with a radius of curvature much larger than the western shell.
At lower surface brightness levels, the arc appears to fill in
toward the west, but cut off more sharly to the east.
The eastern boundary coincides with the dust lanes observed in
the broadband images, as shown in Figures 1c and 2.

The presence of the arc at the inner edge of the dust lanes
suggests that it is ionized by a source near the center of the galaxy,
rather than being simply a part of warm diffuse ionized medium 
which is present in many galaxies 
(Veilleux, Cecil, \& Bland-Hawthorn 1995; Rand 1996).
It is likely that there is a dusty gas disk oriented perpendicular to
outflow axis, and that the top of this gas disk is illuminated by
the central source or the NW shell, causing the upper layers of 
the gas disk to be ionized. The HST optical image of Figure 2 
suggests that the gas disk has significant structure.
It could be that an eastern counterpart exists,
but is not visible because of extinction.

The H$\alpha$/[NII] line ratios in this arc are typically 0.5-0.7,
except for a few compact spots of with H$\alpha$/[NII]$\simeq$1.3
along the northern half of this arc, which may be HII regions.
The line ratio is lower, typically 0.3-0.5, in the fainter gas to the west of arc.
If the arc is ionized by the nuclear source, then the arc
may be similar to ionization cones observed in some Seyfert (NGC~1068 -- 
Macchetto \etal 1994)
and LINER  (NGC~1052 -- Pogge \etal 1999) galaxies,
although it does not have a straight-edge conical shape, has a very large opening
angle, and is relatively faint, distinguishing it from 
what are usually termed ionization cones.

The radial extent of the optical gas and dust disk is
similar to the 13$''$=1 kpc radial extent of CO emission detected in
OVRO interferometer maps (Kenney \etal 1995).
Many of the dust lanes visible near the center of the galaxy
appear to lie in an annular ring at radii of 15$''$, although there are
dust lanes which appear to be non-planar connecting to those 
in the annular ring.

\subsection{Nuclear Source}

The brightest line emission arises from a compact but
resolved source near the center of the galaxy.
This is presumably located at the nucleus,
since it is spatially coincident with 
a bright compact optical continuum source, and 
is within 1$''$ of the centroid 
of the broadband optical isophotes,
which presumably trace the stellar distribution.
A more accurate measure is prevented by 
saturation of the central few pixels in the R and I bands, and by dust.
Both the line source and the B-band continuum source are
resolved with a full width at half power of 0.32$''$=25 pc.

The H$\alpha$/[NII] line ratio of the nuclear source is 1.6$\pm$0.4.
This is similar to many of the HII regions, and higher than the
values in the outflow shell and surrounding diffuse gas.
The central source has an H$\alpha$ luminosity, without any extinction
correction, of 5$\times$10$^{38}$ erg s$^{-1}$,
which is similar to the luminosity of the large HII complex 8$''$ south of nucleus
and an order of magnitude less than 30 Doradus,
but much more compact than either.
If the central H$\alpha$ emission is due to massive star formation,
then the present rate of nuclear massive star formation
is a minimum of 0.004 M$_{\sun }$ yr$^{-1}$, 
since this does not include any extinction correction.

The nucleus appears near the edge of strong dust lanes, 
as shown in Figures 1c and 2.
This implies that the extinction towards the
central few hundred pc of NGC~4438 will be variable,
although the nucleus itself may be viewed through relatively less
extinction than regions immediately to the SE.

The extinction toward the nuclear source can be roughly approximated from 
the H$\alpha$/H$\beta$ ratio from ground-based spectra of Keel (1983)
or Ho \etal (1997). The spectrum of Ho, obtained with a 2$''$$\times$4$''$ aperture,
has an H$\alpha$/H$\beta$ ratio of 5.67, which compares with the expected
ratio of 2.78 for Case B recombination, implying an extinction of
A$_{\rm V}$=3.3 mag, corresponding to 2.6 mag at  H$\alpha$
(Cardelli, Clayton, \& Mathis 1989).
The NW shell has 7 times the H$\alpha$+[NII] luminosity of the nucleus,
so a significant fraction of the flux in ground-based apertures
centered on the nucleus comes from the NW shell.
Thus the extinction estimate of 2.6 mag
is a weighted combination of the extinctions for the nucleus and
the NW shell, and the true nuclear extinction could be greater than 2.6 mag,
both because of the large aperture in which the  H$\beta$ and H$\alpha$
lines are measured, and because extinction estimates from optical
line ratios generally provide a lower limit on the true extinction.
If the extinction at  H$\alpha$ is 2.6 mag, which corresponds to
a factor of 11, then the present rate of nuclear massive star formation
would be 0.05 M$_{\sun }$ yr$^{-1}$.

There is a weak radio continuum counterpart to this optical line source
(Hummel \& Saikia 1991), as shown in Fig 4.
The nuclear flux density estimated from the map of Hummel \& Saikia (1991)
is 1.5$\pm$0.3 mJy at 4.86 GHz,
which corresponds to a power of 5.0$\times$10$^{19}$ W Hz$^{-1}$.
The spectral index of the nucleus is uncertain, since it is faint and 
unresolved in the 1.4 GHz map of Hummel \& Saikia (1991), as shown in Figure 3.
The radio continuum emission at 4.86 GHz from star forming regions in
galaxies is a mixture of thermal and non-themal emission. If the spectral index of 
non-thermal component of the nuclear source has the typical value of 0.8, 
then at 4.86 GHz about 3/4 of the flux is non-thermal, and 1/4 is thermal 
(Condon 1992).
If this radio continuum power is all due to star formation, then
the steady-state total star formation rate is about 0.1 M$_{\sun}$ yr$^{-1}$,
according to the relations given in Condon (1992).
This yields an upper limit 
of a factor of 24, or 3.4 magnitudes,
on the extinction at H$\alpha$ toward the nuclear source, if
the radio continuum emission is dominated by star formation.
Given the uncertainties, 
this agrees well with the  H$\alpha$-based estimate of the nuclear
star formation rate. 

If the nuclear radio continuum and H$\alpha$
emission are powered predominantly by something other than
star formation, 
then the radio continuum emission cannot be used to estimate the
extinction at H$\alpha$, although the upper limit on the
the  nuclear star formation rate would still hold.

\section {Discussion}

\subsection{ Energetics of the Shells}

In order to obtain some kinematic information on the shells,
we have reanalyzed the ground-based optical slit spectra of NGC~4438 
described by Kenney \etal (1995) and Rubin \etal (1999), which was
obtained with the Kitt Peak 4m telescope.
The spectrum taken at a position angle of 104$\deg$,
is the closest to the shell  position angle of 125$\deg$,
and passes through the western edge of the NW shell.
Given the slit orientation and the limited spatial resolution (2$''$)
of this ground-based data,
we are not able
to extract a useful limit on the velocity gradient across the shell, but
can provide some linewidth information.
The FWHMs of the brighter of the [NII] and [SII] lines are 
measured in 2$''$ apertures to be 408 km s$^{-1}$ at the nucleus,
$\sim$200 km s$^{-1}$ at 5-10$''$ from the nucleus (well beyond the shell),
and 290 km s$^{-1}$ at 2$\pm$1$''$ from the nucleus, where the slit
covers the the western edge of the NW shell.
This linewidth would contain contributions from both
bulk and turbulent motions, but could greatly
underestimate the bulk outflow motions.
The true internal velocities could be significantly higher,
since the maximum velocities are likely projected close to the plane of
the sky, and since the largest line-of-sight velocities
are probably occur near the bubble mid-axis (as in NGC~3079, Veilleux \etal 1994),
and our ground-based spectrum misses the mid-axis of NGC~4438's bubble.
We nonetheless adopt a reasonable lower limit on the internal motions
of the brightest optical line emitting gas in the NW shell
of 300 km s$^{-1}$.

The [SII] doublet ratio $\lambda$6716/$\lambda$6731 
at the western edge of the NW shell 
is measured from our Kitt Peak 4m spectrum to be 1.08$\pm$0.01,,
which corresponds to an electron density of 420 cm$^{-3}$,
for an electron temperature of 10$^4$ K (Osterbrock 1989).

The H$\alpha$ luminosity of the NW shell is 3.4$\times$10$^{39}$ erg s$^{-1}$,
assuming no extinction correction,
which corresponds to a mass of ionized gas of  3.5$\times$10$^{4}$ M$_{\sun}$,
assuming an electron density of 420 cm$^{-3}$
(Osterbrock 1989; Veilleux, Shopbell \& Miller 2001).
Assuming the same mass for the SE shell, the total ionized gas mass
of 7$\times$10$^{4}$ M$_{\sun}$ for NGC~4438
would be somewhat less than, although similar to, the 
ionized gas masses in the outflows of M82 (2$\times$10$^{5}$ M$_{\sun}$)
and NGC~3079  (1$\times$10$^{5}$ M$_{\sun}$) (Veilleux \etal 1994).

This ionized gas mass of the NW shell has an associated kinetic energy of
6$\times$10$^{52}$ erg, assuming a velocity of 300 km s$^{-1}$.
Assuming the same KE for the SE shell, 
the KE for the 2 shells would be 1.2$\times$10$^{53}$ erg.
This is a lower limit on the KE, since it does not include atomic, molecular,
or hotter ionized gas components, does not include any extinction correction, 
and is based on a lower limit to the internal velocities.
This KE is similar to the that associated with the ionized gas in the
outflow in M82 (2$\times$10$^{53}$ erg), although significantly lower
than that in  NGC~3079  (2$\times$10$^{54}$ erg) (Veilleux \etal 1994).
The mass and kinetic energy associated with the ionized gas in the
outflow of NGC~4438 is comparable to that in M82, despite the fact
that the central star formation rate in NGC~4438 appears to be
roughly one order of magnitude lower.

\subsection{ A Starburst or AGN Origin?}

What is the nature of the nuclear energy source driving the outflows?

The lower limit on the kinetic energy in the shells of 10$^{53}$ erg
is equivalent to about 100 supernovae.
Assuming a Salpeter initial mass function
and an upper mass cutoff of 100 M$_{\sun }$, 
this would correspond to a total mass of
more than 10$^4$ M$_{\sun }$ in stars formed within the last 10$^6$ yr
(Elson \etal 1989), and an average
star formation rate of 0.01 M$_{\sun }$ yr$^{-1}$. 
Since the base of the outflow seems
to be quite narrow, this activity would need to have occured within the
central 1$''$=77 pc.
The upper limit on the present rate of massive star formation in the nucleus, 
as estimated from 
the extinction-corrected nuclear H$\alpha$ emission 
or the  nuclear radio continuum emission,
is 0.05-0.1 M$_{\sun }$ yr$^{-1}$ (\S 3.4). 

Thus a starburst origin for the shells is energetically plausible,
given the present estimate of the kinetic energy in the shells.
Moreover, the H$\alpha$/N[II] ratio of the nucleus is consistent with that
of an HII region.
Although no known starburst outflow has
the large degree of collimation of the SE shell,
or strongly enhanced radio continuum emission at the outer ends of the shells,
it might be that the outflow from the formation of a compact nuclear star
cluster could be sufficiently collimated to produce the observed
working surfaces.

The outflow axis is shifted by $\sim$6$\deg$ from the apparent minor 
axis of the 100-1000 pc scale CO disk. This is not strong evidence
for either the starburst or AGN scenario.
A large misalignment would indicate an AGN jet, whereas alignment
would be consistent with an extended starburst.
Given the disturbed morphology of the galaxy, we do not attribute
much significance to the relatively small offset.

While some properties of the outflow
are consistent with a nuclear starburst outflow,
others are more indicative of jets associated with nuclear black holes.
The low level of star formation in the central 100 pc, the
collimation of the SE shell, the enhanced radio continuum brightness
at the outer ends of the shells, and the broad line component,
are all suggestive of a nuclear power source more exotic than star formation.

The broad line component with FWHM=2050 km s$^{-1}$ reported by Ho etal (1997)
is suggestive of a ``true AGN'', and might arise from an accretion disk.
However, lines this wide can also occur in outflowing gas.
While most outflows associated with superbubbles have velocities
of a few hundred  km s$^{-1}$ (Heckman, Armus, \& Miley 1990),
the starburst-plus-AGN galaxy NGC~3079 has extranuclear H$\alpha$ velocities of
2000 km s$^{-1}$ in the outflowing gas (Veilleux \etal 1994).
So the existence of a moderately broad component to the  H$\alpha$ line
in NGC~4438 does not by itself distinguish between an AGN and starburst
as the driver of the outflow.

A possible problem with the AGN jet scenario is that no jet has been detected.
Most of the radio continuum emission in Figs. 3 and 4 arises from the edges and
ends of the shells.
While no jet has been detected, there are
suggestions that some Seyfert jets are sub-relativistic and weak emitters of
synchrotron radiation on ~100 pc scales
(Bicknell \etal 1998). Much of the energy in Seyfert jets may be associated with
thermal rather than relativistic plasma. 

Since NGC~4438 has a large bulge with a luminosity of 10$^{10}$ L$_{\sun}$,
a large nuclear black hole might be expected.
There is no published stellar velocity dispersion for NGC~4438,
but the rough correlation between bulge luminosity and black hole mass
shown by  Ferrarese \& Merritt (2000) or Gebhardt \etal (2000)
suggests a black hole mass of $\sim$5$\times$10$^{7}$ M$_{\sun}$ for NGC~4438.
Such a large black hole mass should produce a kinetic signature
detectable with HST resolution.

The nuclear region of NGC~4438 hosts an x-ray source 
detected by both the Einstein and ROSAT satellites.
The Einstein High Resolution Imager
measured a luminosity from 0.1-2 keV
of Lx = 3x10$^{39}$ erg s$^{-1}$
(Kotanyi, van Gorkom \& Ekers 1983),
whereas the ROSAT High Resolution Imager measured
a significantly higher luminosity from 0.1-2.4 keV
of Lx = 1.3$\times$10$^{40}$ erg s$^{-1}$
(Roberts \& Warwick 2000).
The nuclear source is unresolved by both the Einstein and the ROSAT
HRIs, which have resolutions of $\simeq$10$''$,
so it is unclear how much of this emission
arises from the nucleus and how much from the outflows.
Even if all of this emission arises from the nuclear source,
the observed x-ray luminosity (even with the lower Einstein value)
is significantly below
the Eddington luminosity for a black hole mass of 5$\times$10$^{7}$ M$_{\sun}$,
suggesting accretion onto the black hole at a sub-Eddington rate.
This appears to be the situation 
for most nearby galaxies (Roberts \& Warwick 2000),
and is consistent with the accretion rates suggested for
advection-dominated accretion flows (ADAFs -- Narayan \& Yi 1995; 
Fabian \& Rees 1995).

NGC~4438 has also been detected in x-rays by the ASCA satellite, 
which provides information about the x-ray spectrum.
Terashima, Ho, \& Ptak (2000)
found a reasonably good correlation between the hard (2-10 kev) ASCA x-ray
luminosity and the H$\alpha$ luminosity in the central 2$''$$\times$4$''$
for LINER 1 galaxies, 
defined to be those LINERs which contain a broad
component to the  H$\alpha$ line.
The correlation is poor for LINER 2 galaxies.
This provides evidence that the hard x-ray emission from an AGN 
is the source of ionization for the optical line emission observed
near the nuclei of most LINER 1 galaxies, and that some other
mechanism is likely for LINER 2 galaxies.

The only LINER 1 galaxy in their sample which deviated strongly from this
correlation was NGC~4438, which had a ratio of
hard x-ray to H$\alpha$ luminosity  one to two orders of magnitude
lower than the rest.
This could be partly, but not entirely, due to strong H$\alpha$ emission from 
the NW outflow shell
in NGC~4438, which contributes significantly to the flux in a  2$''$$\times$4$''$
aperture. The ionized gas associated with the shell is likely shock excited
due to the outflow.
If one compares the hard x-ray  luminosity with 
H$\alpha$ luminosity from the nuclear source only
(about half the total in the central 2$''$$\times$4$''$),
then NGC~4438 is somewhat closer to the other LINER 1's, but still 
has a significantly lower ratio of L$_{\rm x}$/L$_{\rm H\alpha}$.
This suggests that NGC~4438 is a significantly different nuclear 
source from most LINER 1's, and that the broad H$\alpha$ 
component might be
gas in the outflow, and not a true broad line component
(e.g., emission from an accretion disk).

%\subsection{ Environmental Effects}
\subsection{ The Asymmetries of the Outflow Shells}

\begin{figure}
\epsscale{0.9}
\plotone{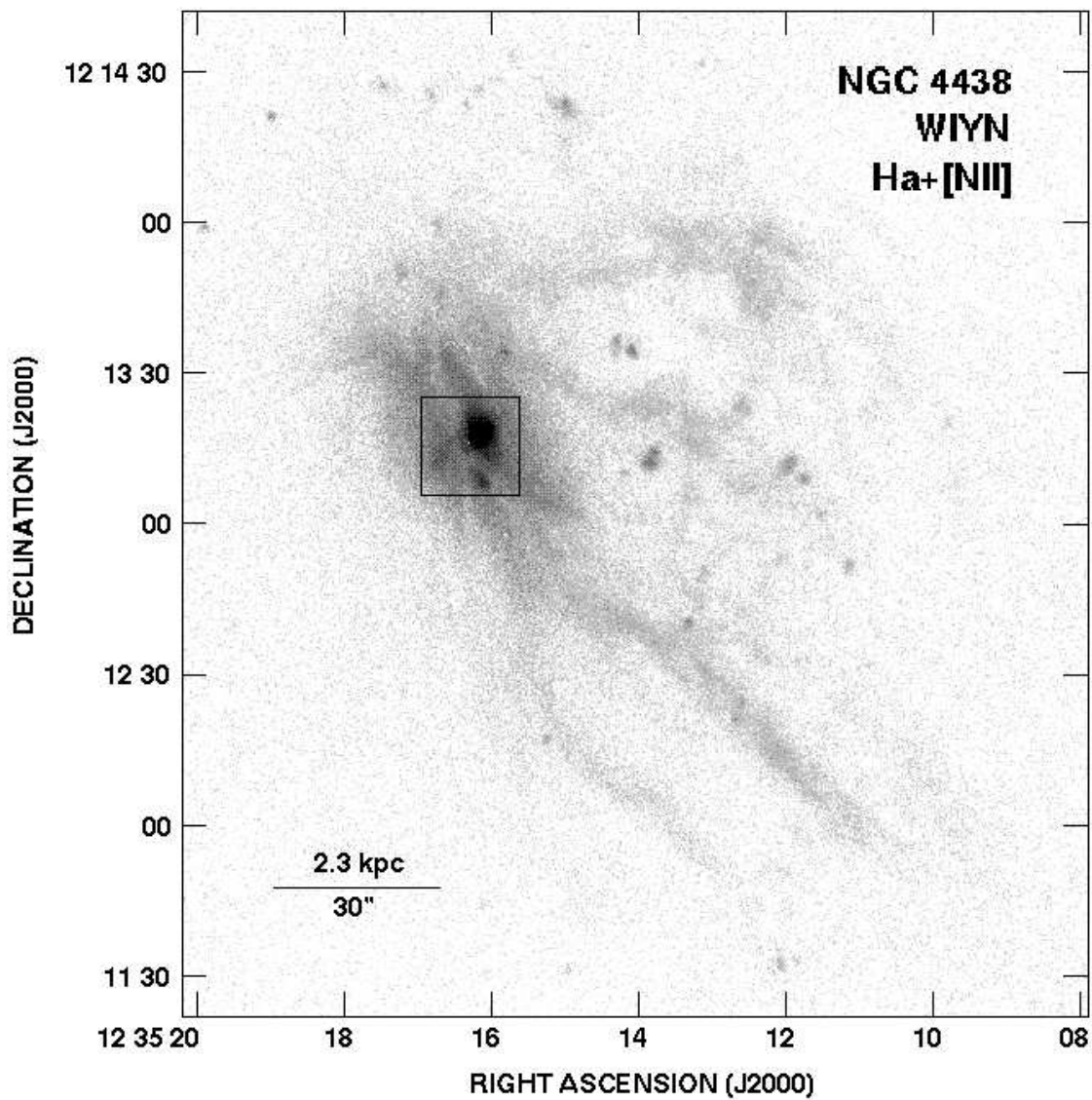}
\caption{
WIYN H$\alpha$+[NII] image of NGC~4438,
with resolution of 1$''$ and 180$''$$\times$200$''$ field of view, 
shown with a logarithmic stretch.
This image shows the larger scale environment into which the nuclear outflow is propogating.
The inset box outlines the region shown in Figures 1 and 3.
The nucleus plus NW shell, the SE shell, the strong dust lane crossing the
SE shell, and the HII complex 8$''$ of
the nucleus can all be seen within the inset region.
In addition, 2 larger-scale ionized gas components are apparent --
a diffuse component of $\sim$30$''$$\times$60$''$ extent,
and a one-sided filamentary nebula extending out 120$''$ to the NW.
There are HII regions located within the one-sided filamentary nebula,
although only a small fraction of the emission arises from HII regions.
Note that the large scale ISM asymmetry in NGC~4438, with emission on the NW side
much stronger than on the SE side, is the same as that observed for the
nuclear outflow shells.
}
\end{figure}

Some properties of the nuclear emission regions
may be due to peculiarities in the circumnuclear environment
caused by the interaction(s) experienced by NGC~4438.
We would not be the first to speculate that an interaction 
has increased the flow of gas into the nucleus, fueling a nuclear 
starburst or feeding a central black hole, although it is difficult
to establish any direct causal connections between nuclear fueling
and the large-scale envirnoment for any one particular galaxy.
On the other hand, a case can be made that the asymmetries in 
the outflow shells are associated with the global disturbance to NGC~4438.

The differences in luminosity, length, and opening angle
between the two shells
are among the most striking features of the nuclear outflow in NGC~4438.
The SE shell is 2.5 times longer, 15 times
less luminous in the radio continuum, and has an opening angle much narrower
than the NW shell.
These 3 differences can all be explained in principle
by jet material running into an ISM which is denser
on the NW side of the disk than the SE.
Even if the energy injected is the same on both sides,
a larger fraction of the energy is dissipated on the NE side
as the outflowing material is significantly decelerated, 
resulting in more radiation.
Similarly, the impedance associated with a higher ISM density 
has been considered at least part of the reason why
jets in spiral galaxies tend to have wider opening angles and smaller lengths
than those in ellipticals  (Elmouttie \etal 1998).
The higher luminosity together with the compact morphology of the NW shell 
suggest
that it is at the stage of superbubble evolution
in which radiative losses are dynamically important
(e.g., Weaver \etal 1977; Ostriker \& McKee 1988).
On the other hand, the SE shell is likely to be in an earlier, 
free-expanding or adiabatic phase.

A systematic difference in the circumnuclear ISM density 
on the 2 sides of the disk 
seems plausible, since the ISM of NGC~4438 shows this asymmetry on 
larger scales.
Much of the HI, CO, H$\alpha$, radio continuum, and x-ray emission
are strongly displaced to the west of the stellar disk
(Kotanyi \etal 1983; Combes \etal 1988; Cayatte \etal 1990; 
Kenney \etal 1995), presumably due to an interaction in the cluster.
Figure 5 shows a ground-based H$\alpha$+[NII] image from the WIYN telescope,
over a 180$''$$\times$200$''$ field of view, with 1$''$ resolution
(Kenney \etal in prep). This shows the ISM environment into which
the nuclear outflow is propogating.
Inside the box in the center, which outlines the 18$''$ field of view shown in
Figures 1 and 3, the nucleus plus NW shell, the SE shell, the dust lane 
crossing the SE shell, and the HII complex 
8$''$ of the nucleus can all be seen. 
In addition, 2 larger-scale ionized gas components are apparent --
a diffuse component of $\sim$30$''$$\times$60$''$ extent,
and a one-sided filamentary nebula extending out 120$''$ to the NW.
There are HII regions located within the one-sided filamentary nebula,
although only a small fraction of the emission arises from HII regions.
Note that the large scale ISM asymmetry in NGC~4438, with emission on the NW side
much stronger than on the SE side, is the same as that observed for the
nuclear outflow shells.

If this same asymmetry extends down to scales of ~10 pc,
this could explain why the NW shell is so much brighter and shorter
than the SE shell.
The gas, which may be slowly infalling back into NGC~4438 
after a galaxy-galaxy collision (Kenney \etal 1995),
may not fall much further than the disk midplane, 
since it will collide with the gas already in the disk.
The strong emission of the NW shell may arise 
because the ISM above the disk plane is unusually dense,
due to the disturbance experienced by NGC~4438,
resulting in a stronger interaction between the outflow and the ambient ISM.

We have examined the [SII]  $\lambda$6716/$\lambda$6731 doublet ratio
from the ground-based optical spectrum cited above (along PA=104$\deg$ )
in order to see whether there is any difference in ionized gas density
of the 2 sides of the galaxy. 
The [SII] ratios measured 6$\pm$1$''$ from the nucleus on both sides of
the galaxy do show a significant difference.
On the NW side, a few arcseconds beyond the NW shell, the [SII] ratio is
1.07$\pm$0.02, which corresponds to an electron density of 450 cm$^{-3}$,
whereas at the same distance from the galaxy center toward the SE,
the [SII] ratio is
1.15$\pm$0.02, which corresponds to a lower electron density of 310 cm$^{-3}$.
This provides evidence that the gas density on the 2 sides
of the galaxy disk may indeed be different.

\subsection{Other Galaxies with  Similar Outflow Morphologies}

Many active galaxies contain one-sided or bipolar nuclear narrow line emission features, 
including jets, shells, bubbles or ionization cones, which extend 100's to 1000's of pcs
(e.g., Wilson \etal 1993; Wilson \& Tsvetanov 1994; Macchetto \etal 1994;
Baum \& Heckman 1989; Baum \etal 1993; Lenhert \& Heckman 1996; Falcke \etal 1998).
The morphologies vary widely, even among galaxies with a given
nuclear classification.
For example, Pogge \etal (1999) have recently imaged the line emission 
in 14 LINERs with HST, and found a variety of morphologies,
few of them well-ordered, and none of them much like NGC~4438.
Those galaxies whose nuclear  regions
most closely resemble the one in NGC~4438
are those with closed shells, double bubble, or figure-8 shaped 
morphologies, in narrow emission lines and/or the radio continuum.
There are such features in both Seyfert and starburst/LINER galaxies,
and it is not always clear whether they are formed by AGN jets
or starburst winds since both can form bubbles.

The Seyfert 2 galaxy NGC~2992  
exhibits a figure-8 shaped morphology in the radio continuum
(Ulvestad \& Wilson 1984), although the narrow line morphology
is dominated by an ionization bi-cone apparently unrelated
to the radio continuum lobes (Allen \etal 1999).
The ionization from the strong Seyfert nucleus of this galaxy
may overwhelm that from the edge-brightened bubbles
of radio plasma producing the radio lobes.
Although nuclear jets may be responsible for blowing the nuclear bubbles,
there is no strong enhancement in radio emission at the outer ends of the lobes,
and therefore no evidence of working surfaces 
in NGC~2992, in contrast to both NGC~4438 and the Seyfert 2 galaxy NGC~3393.
NGC~3393 contains a narrow-line region dominated by
S-shaped arms which border a linear, triple-lobed radio source 
(Cooke \etal 2000). The origin of the S-shaped emission line region 
is not clear, but might
result from either an incomplete bipolar bubble or precessing jets.
The strongest radio emission associated with the optical arms
does not trace the optical arms,
but arises from smaller regions which are spatially coincident
with part of the optical arms. These are presumably working surfaces
where a jet is strongly interacting with the surrounding ISM.

The starburst/LINER galaxies NGC~2782 (Jogee \etal 1998, 1999)
and NGC~3079 (Ford \etal 1986; Veilleux \etal 1994; Cecil \etal 1999;
Veilleux 2000) exhibit partially closed shell morphologies in 
emission lines and the radio continuum.
The optical and radio shells are roughly spatially coincident
in NGC~2782 (Jogee \etal 1998) whereas in NGC~3079 the radio shell is signficantly more
extended than the optical shell (Ford \etal 1986). 
This suggests that  NGC~2782 may be in a pre-blowout phase of starburst 
superwind evolution (Chevalier \& Clegg 1985; Tomisaka \& Ikeuchi 1988; 
Heckman \etal 1990), and that NGC~3079 may be in an early blowout phase. 
While the role of AGN in these 2 galaxies cannot be discounted, both of 
these galaxies contain nuclear starbursts, and neither exhibits 
a strong enhancement in radio continuum at the outer ends of their lobes.
These properties distinguish them from NGC~4438.

Further exploration into the differences between these 
galaxies may be helpful in general for distinguishing
between the outflows of nuclear star formation and supermassive black holes.

\section{Future Work}

Ultimately we wish to know nature of the nuclear source and
whether the outflow is driven by AGN or starburst.
The present study has not given definite answers for these questions,
and future studies will help provide the answers.

High spatial resolution optical or infrared spectroscopy of the nucleus
could show the spatial distribution of the 
2000 km s$^{-1}$ broad line component, revealing whether it
originates in the outflowing gas or is a true AGN broad line component.
It could also show
whether there is a very broad line component on sub-arcsecond scales
possibly arising from
an accretion disk, as in the Virgo LINER NGC~4579 (Barth \etal 2001).
Studies of the stellar kinematics from absorption lines
can constrain the mass of a central black hole and show whether
the outflow originates from the dynamical center of the galaxy.
IR imaging and spectroscopy can provide
better extinction estimates, further revealing the true nature of the nucleus,
and yielding new information on the spatial extent and intensity of the
central star formation activity.

High resolution x-ray observations (Chandra) can show how much of the
central 10$''$ Einstein and ROSAT x-ray source is associated with the nucleus,
and how much with the outflow.
The x-ray spectrum of the nuclear source can
reveal the true nature of the nucleus, 
whereas that of the outflow can reveal whether the hot gas 
has sufficient energy to be
dynamically important for the  evolution of the bubble.
In M82 the high temperature (4$\times$10$^7$ K)
thermal x-rays may contain enough pressure to drive the
outflow (Griffiths \etal 2000).

High resolution molecular gas (e.g. CO) observations 
may be important for learning the energy budget of bubbles,
since molecular gas may be participating in the outflow,
and could be carrying a significant amount of kinetic energy.
Most of the kinetic energy associated with a starburst superbubble in M82
appears to be in the form of molecular rather than ionized
gas (Matsushita \etal 2000),
and it will be interesting to learn whether the bubbles in NGC~4438 are
similar.

We find it remarkable that there is a such a striking outflow
from a nucleus without a very luminous starburst or AGN,
and think followup studies of NGC~4438 and similar galaxies
are important because of the possible
implications for the deposition of matter and energy 
into galactic haloes and the intergalactic medium.

\section{ Acknowledgements}

We thank the referee Sylvain Veilleux for helpful comments
which resulted in an improved paper, 
and Simon Cassasus for analyzing the ground-based spectra,
and for his interest.
We are also grateful to Andrew Baker, Stefi Baum, Paolo Coppi, Carol Mundell,
Chris O'Dea, Megan Urry, 
and Andrew Wilson for helpful exchanges, and
Zolt Levay for producing the beautiful color images.
This work has been supported by Space Telescope grant GO-06791.

\eject


\begin{references}

\reference{} Allen, M. G., Dopita, M. A., Tsvetanov, Z. I., \& Sutherland, R. S.
 1999, ApJ 511, 686
\reference{} Barth, A. J., Ho, L. C., Filippenko, A. V., Rix, H.-W., \&
 Sargent, W. L. W. 2001, ApJ, 546, 205
\reference{}  Baum, S. A., \& Heckman 1989, ApJ, 336, 702
\reference{}  Baum, S. A., O'Dea, C. P., Dallacassa, D., de Bruyn, A. G., \& 
 Pedlar, A. 1993, ApJ, 419, 553
\reference{} Bicknell, G. V., Dopita, M. A., Tsetanov, Z. I., \& Sutherland, R. S. 1998,
  ApJ, 495, 680
\reference{} Capetti, A., Axon, D. J., Macchetto, F. D., Marconi, A., \& Winge, C. 1999,
 ApJ, 516, 187
\reference{} Cardelli, J. A., Clayton, G. C., \& Mathis, J. S. 1989, ApJ, 345, 245
\reference{} Cayatte, V., van Gorkom, J. H., Balkowski, C., \& Kotanyi, C.
1990, AJ, 100, 64
\reference{} Cecil, G. 1988, ApJ, 329, 38
\reference{} Cecil, G., Veilleux, S., Filippenko, A. V., \& Bland-Hawthorn, J.
1999, BAAS, 195, 0808C
\reference{} Combes, F., Dupraz, C., Casoli, F., \& Pagani, L. 1988, AA, 203, L9
\reference{} Condon, J. J. 1992, ARAA, 30, 575 
\reference{} Cooke, A. J., Baldwin, J. A., Ferland, G. J., Netzer, H., \& Wilson, A. S.
   2000, ApJS, 129, 517
\reference{} Elmouttie, M., Haynes, R. F., Jones, K. L., Sadler, E. M., \& Ehle, M.
  1998, MNRAS, 297, 1202
\reference{} Elson, R. A., Fall, S. M., \& Freeman, K. C. 1989, 336, 734
\reference{} Fabian, A. C., \& Rees, M. J. 1995, MNRAS, 277, 55
\reference{} Falcke, H., Wilson, A. S., \& Simpson, C. 1998, ApJ, 502, 199
\reference{} Ferrarese, L., \& Merritt, D. 2000, ApJ, 539, L9
\reference{} Filippenko, A. V. 1996, in the Physics of LINERs in View of Recent 
  Observations, ed. M. Eracleous et al. (San Francisco: ASP), 17
\reference{} Ford, H. C., Dahari, O., Jacoby, G. H.,
 Crane, P. C.,  \& Ciardullo, R. 1986, ApJ, 311, L7
\reference{} Freedman, W. L. \etal 1994, Nature, 371, 757
\reference{} Gebhardt, K.  \etal 2000, ApJ, 539, L13
\reference{} Heckman, T. 1980 AA, 87, 152
\reference{} Heckman, T. M., Armus, L., \& Miley, G. K. 1990, ApJS, 574,
833
\reference{} Heckman, T. M., Baum, S. A., van Breugel, W. J. M., \&
McCarthy, P. 1989,
	ApJ, 338, 48
\reference{} Ho, L. C. 1999, ApJ, 510, 631
\reference{} Ho, L. C., Filippenko, A. V., \& Sargent, W. L. W. 1997,
	ApJS, 112, 315
\reference{} Hummel, E., \& Saikia, D. J., 1991, AA, 249, 43
\reference{} Jacoby, G. H. \etal 1992, PASP, 104, 599
\reference{} Jogee, S., Kenney, J. D. P., \& Smith, B. J. 1998, ApJL, 494, L185.
\reference{} Jogee, S., Kenney, J. D. P., \& Smith, B. J. 1999, ApJ, 526, 665. 
\reference{}  Katsiyannis, A. C., Kemp, S. N., Berry, D. S., \& Meaburn, J. 1998,
  AAS, 133, 397
\reference{} Keel, W. C. 1983, ApJ, 269, 466
\reference{} Keel, W. C., \& Wehrle, A. E. 1993, AJ, 106, 236
\reference{} Kenney, J. D. P., Rubin, V. C., Planesas, P., \& Young, J. S.
1995, ApJ, 438, 135
\reference{} Kennicutt, R. C., Jr., 1983, ApJ, 272, 54.
\reference{} Kennicutt, R. C., Jr., Edgar, B. K., \& Hodge, P. W. 1989,
	ApJ, 337, 761.
\reference{} Kormendy, J., \& Richstone, D., 1995 ARAA, 38, 581
\reference{} Kotanyi, C. G., van Gorkom, J. H., \& Ekers, R. 1983, ApJ,
273, L7
\reference{} Lenhert, M. D., \& Heckman, T. M. 1996, ApJ 462, 651
\reference{} Macchetto, F., Capetti, A., Sparks, W. B., Axon, D. J., \& Boksenberg, A.
 1994, ApJ, 435, L15
\reference{} Magorrian, J. \etal 1998, AJ, 115, 2285.
\reference{} Moore, B., Katz, N., Lake, G., Dressler, A., \& 
	Oemler, A., Jr. 1996, Nature, 379, 613
\reference{} Narayan, R., \& Yi, I.  1995, ApJ, 452, 710
\reference{} Osterbrock, D. E., 1989, Astrophysics of Gaseous Nebulae and Active Galactic Nuclei,
 (Mill Valley, CA: University Science Books)
\reference{} Ostriker, D. E., \& McKee, C. F. 1988, Rev. Mod. Phys., 60, 1
\reference{} Pogge, R. W., Maoz, D., Ho, L. C., \& Eracleous, M. 
2000, ApJ, 532, 323
\reference{} Rand, R. J. 1996, ApJ, 462, 712
\reference{} Roberts, T. P., \&  and Warwick, R. S. 2000, MNRAS, 315, 98
\reference{} Rubin, V. C., Waterman, A. W., \& Kenney, J. D. P. 1999, AJ, 118, 236.
\reference{} Stauffer, J. R. 1982, ApJ, 262, 66 
\reference{} Terashima, Y., Ho, L. C., \& Ptak, A. F. 2000, 539, 161
\reference{} Tomisaka, K., \& Ikeuchi, S. 1988, ApJ, 330, 695.
\reference{} Ulvestad, J. S., \& Wilson, A. S. 1984, ApJ, 285, 439.
\reference{} Ulvestad, J. S., Wilson, A. S., \& Sramek, R. A. 1981, ApJ, 247, 419.
\reference{} Veilleux, S., 2000, ASP Conf. Series, Vol. 195, ed.
  W. van Breugel \& J. Bland-Hawthorn, 277
\reference{} Veilleux, S., Cecil, G., Bland-Hawthorn, J., Tully, R. B., Filippenko, A. V., 
 \& Sargent, W. L. W., 1994, ApJ 433, 48
\reference{}  Veilleux, S., Shopbell, P. L., \& Miller, S. T. 2001, AJ, 121, 198
\reference{}  Veilleux, S., Cecil, G., \& Bland-Hawthorn, J., 1995, ApJ, 445, 152
\reference{} Weaver, R., \etal 1977, ApJ, 218, 377
\reference{} Wilson, A. S., \& Tsvetanov, Z. I. 1994, AJ, 107, 1227
\reference{} Wilson, A. S., Braatz, J. A., Heckman, T. H., Krolik, J. H., \&
 Miley, G. K. 1993, ApJ, 419, L61
\end{references}
\end {document}